\documentclass[preprint2]{aastex}

\slugcomment{draft updated 1/30/06}

\shorttitle{Quiet MXRBs}
\shortauthors{McSwain et al.}

\begin{document}

\title{A Spectroscopic Study of Field and Runaway OB Stars}

\author{M.\ Virginia McSwain\altaffilmark{1,2}}
\affil{Department of Astronomy, Yale
University, P.O.\ Box 208101, New Haven, CT 06520-8101}
\email{mcswain@astro.yale.edu}

\author{Tabetha S.\ Boyajian\altaffilmark{1}, Erika D.\
Grundstrom\altaffilmark{1}, Douglas R.\ Gies}
\affil{Department of Physics and Astronomy, Georgia State University,
P.O.\ Box 4106, Atlanta, GA 30302-4106}
\email{tabetha@chara.gsu.edu, erika@chara.gsu.edu, gies@chara.gsu.edu}

\altaffiltext{1}{Visiting Astronomer, Kitt Peak National Observatory, 
National Optical Astronomy Observatories, operated by the Association of 
Universities for Research in Astronomy, Inc., under contract with the 
National Science Foundation (NSF).}
\altaffiltext{2}{NSF Astronomy and Astrophysics Postdoctoral Fellow}


\def\kms    {\ifmmode{{\rm km~s}^{-1}}\else{km~s$^{-1}$}\fi}
\def\Mdot   {\ifmmode {\dot M} \else $\dot M$\fi}
\def\Mspy   {\ifmmode {M_{\odot} {\rm yr}^{-1}} \else 
$M_{\odot}$~yr$^{-1}$\fi}
\def\Msun   {$M_{\odot}$}
\def\mum     {\ifmmode{\mu{\rm m}}\else{$\mu{\rm m}$}\fi}
\def\Rstar  {$R_{\star}$}


\begin{abstract}

Identifying binaries among runaway O- and B-type stars offers valuable
insight into the evolution of open clusters and close binary stars.  Here
we present a spectroscopic investigation of 12 known or suspected binaries
among field and runaway OB stars.  We find new orbital solutions for five
single-lined spectroscopic binaries (HD 1976, HD 14633, HD 15137, HD 37737,
and HD 52533), and we classify two stars thought to be binaries (HD 30614
and HD 188001) as single stars.  In addition, we reinvestigate their
runaway status using our new radial velocity data with the UCAC2 proper
motion catalogs.  Seven stars in our study appear to have been ejected
from their birthplaces, and at least three of these runaways are
spectroscopic binaries and are of great interest for future study.

\end{abstract}



\section{Introduction}

Nearly all O- and B-type stars are believed to form in open clusters and
stellar associations, but a handful of OB stars are observed at high
galactic latitudes and with large peculiar space velocities, suggesting
that they have been ejected from the cluster of their birth.  There are
two accepted mechanisms to explain the origin of these runaway O- and
B-type stars.  In one scenario, close multi-body interactions in a dense
cluster environment cause one or more stars to be scattered out of the
region \citep*{poveda1967}.  In rare cases, the ejected stars may be bound
as a binary pair; \citet{leonard1990} use $N$-body simulations of open
clusters to show that a binary frequency of about 10\% is expected from
dynamical ejections.  An alternative mechanism involves a supernova
explosion within a close binary, ejecting the secondary due to the
conservation of momentum \citep{zwicky1957, blaauw1961}.  The resulting
neutron star may remain bound to the secondary if not enough mass is lost
during the explosion.  \citet{portegieszwart2000} predicts a binary
fraction of $20-40$\% among runaways which are ejected by such a binary
supernova scenario.  The observed fraction of binaries among runaways
seems consistent with either scenario ($5-26$\%; \citealt{mason1998}).

Observational evidence suggests that both dynamical interactions and
supernovae in close binaries do produce some runaway systems.  The
microquasar LS 5039 was likely ejected from the Milky Way disk about 1.1
Myr ago by the supernova explosion of the neutron star progenitor
\citep{ribo2002, mcswain2004, casares2005}.  \citet{berger2001} find that
about 7\% of Be stars have high peculiar space velocities, likely due to
energy received during the supernova of a close binary companion
\citep{mcswain2005}. On the other hand, \citet*{gualandris2004} show that
a four-body encounter 2.5 Myr ago in the Trapezium cluster resulted in
three ejected systems: two single stars and one binary.  The single stars
both have very large space velocities and are classified as runaways,
although the ejected binary has a lower space velocity due to its higher
total mass.  In a study of 44 trapezium-type clusters, \citet*{allen2006}
find 4 stars with large transverse velocities that were likely ejected by
dynamical interactions.

Which process dominates the production of runaway stars?  
\citet*{hoogerwerf2001} show that massive runaways tend to have enriched
helium abundances and higher rotational velocities, while fewer
non-runaways exhibit these traits.  While rotational mixing could explain
these results, it is doubtful that mixing occurs preferentially among
runaways.  It is more likely that prior mass transfer of CNO-processed
material from the evolved donors has altered the surface abundances and
added angular momentum to the mass gainers.  The high binding energy of
close massive binaries makes it unlikely they will be disrupted by
dynamical interactions, but they could be separated by a supernova.  Thus
according to \citet{hoogerwerf2001}, the evidence favors the supernova
ejection scenario over dynamical cluster interactions.  However, pulsar
searches by \citet{philp1996} and \citet*{sayer1996} have not identified
any neutron stars among runaway OB stars, and further investigation of
their properties is needed.

The production of runaway OB binaries is expected to be rare, but these
systems can offer key insights into the evolution of close binary stars
and open clusters.  If massive stars form in dense regions through the
mergers of low mass protostellar cores, the expected multi-body
interactions will result in a large number of high velocity runaways
\citep{bally2005}.  Ejecting a star dynamically from a young cluster will
reduce the total energy imparted to the cluster via stellar winds,
radiation, and supernova ejecta, whereas a proto-supernova system will
remain in the cluster for a longer time and contribute more energy into
its environment \citep{dray2005}.  Finally, the relative importance of the
supernova ejection scenario places limits on the mass needed to form black
holes and neutron stars \citep{dray2005}.

In this work, we examine the binary status of 12 field and runaway OB
stars.  Through a radial velocity study of these stars, we determine the
orbital solutions of five single-lined spectroscopic binaries (SB1s).  One
is a possible triple star, and we also find two suspected SB1 systems.  
By fitting the spectral energy distributions of these stars, we measure
their reddening and angular sizes to constrain their distances.  Finally,
our radial velocities combined with proper motions allow us to measure
their peculiar space velocities and reinvestigate the runaway status of
these targets.  Our results reveal that only 7 of the 12 stars have high
space velocities, while five stars may not be runaways.  Of the
runaways, two are SB1s and a third is a suspected SB1.  We will
investigate their cluster ejection scenarios in a forthcoming paper.


\section{Target Selection \label{targets}}

Most of our target stars (HD 14633, HD 15137, HD 30614, HD 37737, HD
52266, HD 52533, HD 60848, HD 188001, and HD 195592) have been classified
as field or runaway stars by \citet{gies1986}, \citet{mason1998}, and/or
\citet{stickland2001}.  Five of these (HD 30614, HD 52266, HD 60848, HD
188001, and HD 195592) have unknown multiplicity or are suspected
single-line spectroscopic binaries (SB1s).  The others are known SB1
systems.  Of this set, HD 14633 and HD 15137 are of particular interest
since both SB1 systems were likely ejected from the open cluster NGC 654,
and their travel times appear longer than the expected O star lifetimes
\citep{boyajian2005}.  On the other hand, the field stars HD 52266, HD
52533, and HD 195592 may be the dominant members of previously undetected
clusters \citep{dewit2004}.

HD 1976 has been classified as a member of the Cas-Tau OB association
\citep{dezeeuw1999} and is a visual binary \citep{docobo1986}.  It is also
a slowly pulsating B star \citep{mathias2001} and a known SB1
\citep{blaauw1963}.  Although it is not a field star or a runaway, we
include it in our observational program because its poorly known orbital
solution \citep{blaauw1963, abt1990} indicates a moderate eccentricity and
a low mass function.  Refining its orbit could help determine if this
system is a post-supernova system whose kick velocity was too small to
eject it from the association.

The star HD 36576 is a field Be star \citep*{zorec2005} and a radial
velocity variable \citep{gies1986}.  While \citet{gies1986} classify this
star as a non-runaway, \citet{berger2001} did not include this star in
their sample when they performed a new investigation of high-velocity Be
stars using \textit{Hipparcos} proper motions.  Therefore we reinvestigate
its runaway status here.

Feige 25 (HIP 12320) is a B-type star at high Galactic latitude ($-48.36^
\circ$) whose abundances and kinematics are consistent with a runaway
Population I star \citep{martin2004, martin2006}.  Its radial velocity has
been measured only twice before by \citet{martin2006} and by
\citet{greenstein1974}, and those measurements differ by nearly 30 \kms.  
Therefore we suspect that it may be an SB1 system, and we include it here
for further investigation.

Most of the selected candidates are not known X-ray emitters, although
three (HD 30614, HD 52533, and HD 188001) are known weak X-ray sources.  
\citet*{meurs2005} showed that the observed X-ray luminosities and
hardness ratios of HD 30614 and HD 188001 are consistent with the emission
of normal O supergiant stars rather than from X-ray binaries.  The
hardness ratios of HD 52533 \citep{voges2000} are likewise consistent with
an origin in a normal O star \citep{motch1998}.  However, 
\citet{meurs2005} point out that these results do not rule out the
presence of a compact companion.  If the accretion rate is sufficiently
low, the resulting X-ray production may not contribute significantly to
the total X-ray luminosity.  A deeper X-ray observation may be able to
detect the characteristic hard spectrum of a MXRB in these systems.

Likewise, none of our targets contain a known pulsar.  \citet{philp1996}
searched for pulsars in 44 OB runaway stars, including our targets HD
30614, HD 37737, and HD 52533.  Similarly, \citet{sayer1996} searched for
pulsars in HD 14633, HD 36576, and 38 other OB runaways.  Although neither
group found any pulsars associated with these stars, their existence
cannot be ruled out.  Using binary population synthesis calculations,
\citet{portegieszwart2000} found that 20\% $-$ 40\% of OB runaways may
have neutron star companions.  However, due to the short lifetime of the
pulsar (assumed to be 10 Myr) and the absorption of the radio emission by
the stellar winds (greater at periastron than apastron),
\citet{portegieszwart2000} predicts that only 1\% $-$ 2\% of these neutron
stars could be visible as radio pulsars at some time during their orbit.


\section{Observations and Radial Velocities}

Each target was observed during two observing campaigns at the KPNO 2.1~m
telescope during 2005 October and November.  We used the \#47 grating (831
grooves mm$^{-1}$)  in 2$^{nd}$ order to obtain a resolving power R =
$\lambda/\delta\lambda \sim 3000$.  The observed wavelength range was
4050$-$4950 \AA, a region that includes numerous H-Balmer, He I, and He II
lines in O- and B-type stars.  The data were reduced using standard
routines in IRAF.

We also obtained some spectra from the KPNO 0.9 m coud\'e feed telescope
in 2005 November.  The resolving power is R $\sim$ 12000 using the long
collimator, grating A (in second order with order sorting filter 4-96),
camera 5, and the T2KB CCD, a 2048 $\times$ 2048 device.  We obtained a
spectral coverage of 4240$-$4580 \AA.

We used all available strong, unblended lines to measure radial velocities,
$V_r$, in these stars.  Rest wavelengths for each line were taken from the
NIST Atomic Spectra Database\footnote{The NIST Atomic Spectra Database is
available online at http://physics.nist.gov/PhysRefData/ASD/index.html.}.  
We selected one spectrum with good signal-to-noise (S/N) to serve as a
template, and we fit the core of each absorption line with a parabola to
determine its absolute radial velocity.  The remaining spectra were then
cross correlated with this template to determine the relative velocities.  
Several lines in the 2.1~m data sets lie near bad pixels which corrupted
the cross correlations, so these lines were usually omitted.  In a few
cases we found systematic differences between sets of lines, and we discuss
these below.  Only in the case of HD 1976, we found a systematic
disagreement between our measurements and those in the literature.  Our
$V_r$ measurements of all other stars were in good agreement with previous
measurements.  For each star, we list the complete set of lines used in
Table \ref{lines}, and the resulting mean velocities are listed in Tables
\ref{vel1} and \ref{vel2}.  For the stars in Table \ref{vel2}, we also
include the standard deviation, $\sigma$, of the $V_r$ measured from many
lines for each observation.

\placetable{lines}

To test the significance of systematic night-to-night (NN) and line-to-line
(LL) variations, we performed a 2-way analysis of variance (2AOV)
statistical test for each set of velocities.  This test is described in
detail by \citet{gies1986}.  Essentially, it allows us to test the null
hypothesis that the star does not exhibit statistically significant
variations.  The $F$ statistic describes the variance of the entire data
set divided by the average variance of each measurement, and $p$ gives the
probability that the observed variations are drawn from the same random
distribution.  If $p$ is small, less than 1\%, then we must reject the null
hypothesis.  Thus a low $p$ indicates that the observed differences in
$V_r$ are significant.  Both $F$ and $p$ depend upon the number of degrees
of freedom, which is just the number of input variables (lines or spectra)
minus one.

The coud\'{e} feed spectral range is limited and includes only a few of the
lines measured in the 2.1~m spectra.  Consequently, we applied the 2AOV
test to the common line sample in 2.1~m data alone, and these results are
summarized in Table \ref{2aov}.  The test indicates significant LL
variations in HD 1976, HD 15137, HD 30614, HD 37737, HD 52533, and HD
60848.  We are not surprised to find LL variations in OB-type stellar
spectra because line formation may occur at different levels in an
expanding atmosphere and because some lines suffer from blending with
weaker nearby components. Furthermore, HD 52533 may have an optical
companion causing contamination in its line profiles.  Even though these
stars do have statistically significant LL variations, we can still
identify significant nightly variations in $V_r$ since we measure a
consistent set of lines in the our spectra.  Feige 25 and HD 60848 show
little evidence of nightly systematic differences, while the nightly
variations observed in all other targets are highly significant.

\placetable{2aov}

For the stars with significant night-to-night $V_r$ variations, we
combined data from the literature (where available) with our own
measurements for improved orbital solutions.  We performed period searches
using a version of the discrete Fourier transform and CLEAN deconvolution
algorithm of \citet*{roberts1987} (written in IDL\footnote{IDL is a
registered trademark of Research Systems, Inc.} by A.\ W.\ Fullerton).  
We then tested any significant signal found in the CLEANed power spectrum
to verify that it produced a reasonable phased $V_r$ curve.  We ruled out
periods less than one day because we never observed rapid, high-amplitude
changes in $V_r$ among our targets.  The best period from the CLEANed
power spectrum was used with the non-linear, least-squares fitting program
of \citet{morbey1974} to solve for the orbital elements.

\begin{figure}
\includegraphics[angle=0,scale=0.8]{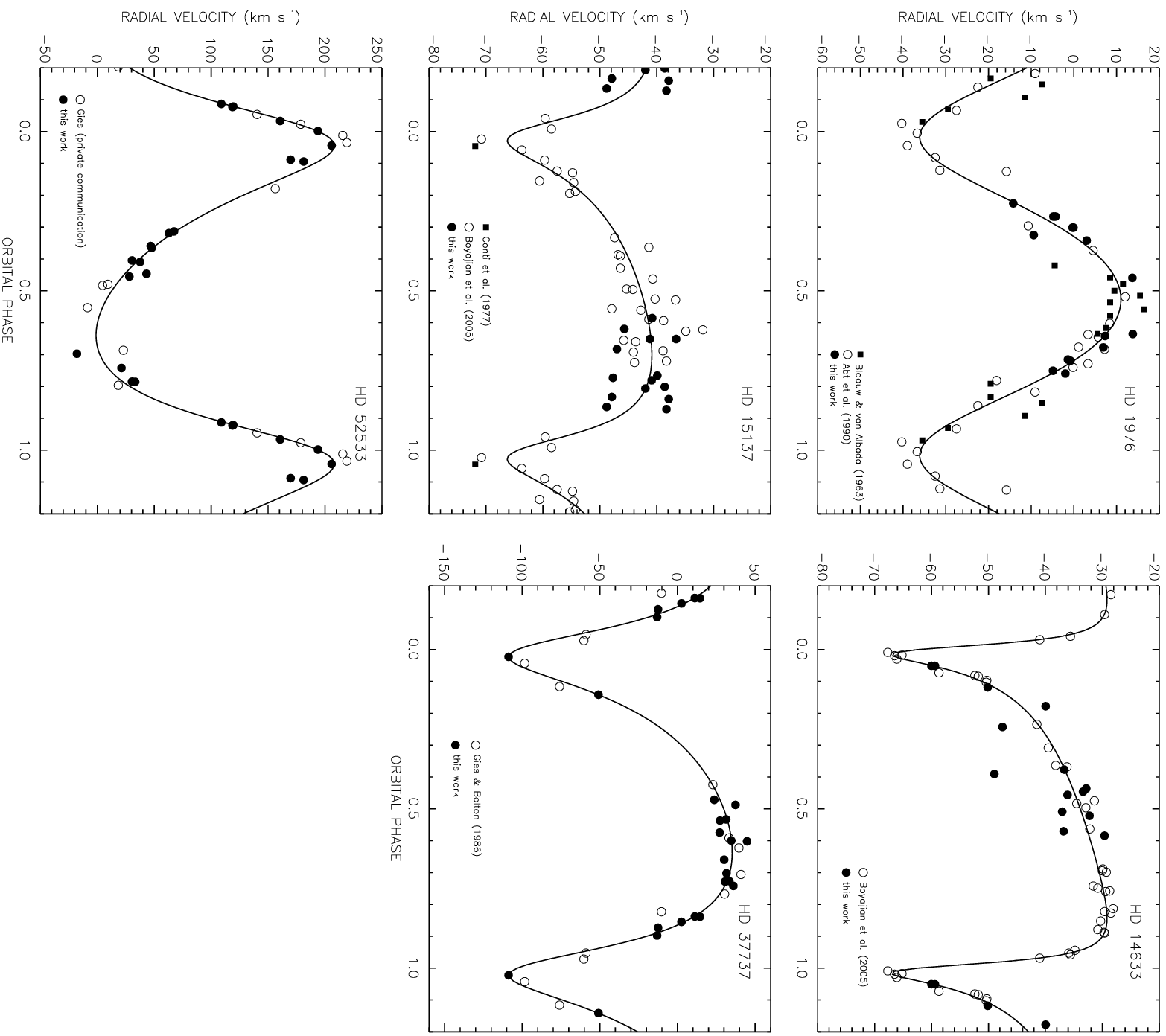}
\caption{
\label{vrmulti}
Radial velocity curves for binaries with orbital solutions.
}
\end{figure}

A detailed discussion of the results from our radial velocity study is
presented below, and we summarize the results here.  We were able to
improve the known orbital elements for 5 SB1 systems (HD 1976, HD 14633,
HD 15137, HD 37737, and HD 52533), and these are listed in Table
\ref{orbits}.  Their $V_r$ curves are shown in Figure \ref{vrmulti}, and
our $V_r$ measurements are listed in Table \ref{vel1}.  The deviations
from the theoretical $V_r$ curves are comparable to the errors of the 
mean.

\placetable{orbits}
\placetable{vel1}

\placefigure{vrmulti}

The stars Feige 25, HD 36576, and HD 60848 appear to be constant velocity
stars.  Both HD 36576 and HD 60848 are Be-type objects, so their $V_r$ are
affected by variable emission filling the line profiles or photospheric
nonradial pulsations changing their line shapes. HD 30614 and HD 188001
have been identified as SB1 systems in the past, but we use available data
in the literature to show that they are more likely single stars that have
been misclassified.  HD 52266 and HD 195592 are likely SB1 systems, but we
cannot determine the systems' periods or orbital elements from the
available data.  The $V_r$ measurements for these stars without orbital
solutions in this work are listed in Table \ref{vel2}.

\placetable{vel2}

\subsection{HD 1976}

HD 1976 is a speckle binary with $\delta V$ = 0.89 and separation
$0\farcs138$ \citep{hartkopf2000, mason2001}.  It is also an SB1, and
\citet{blaauw1963} and \citet{abt1990} each present orbital solutions for
HD 1976.  These solutions have somewhat differing orbital period (27.8 d
and 25.44 d, respectively) and eccentricity (0.2 and 0.14, respectively).

Unfortunately, our observing schedule did not allow us to obtain good
orbital coverage of this system, and we rely heavily upon the data of
\citet{blaauw1963} and \citet{abt1990} to determine the period and orbital
elements.  To update the orbital elements for this SB1, we included our
own 15 $V_r$ measurements with those published by \citet{blaauw1963} and
\citet{abt1990} (except 4 points with large scatter, which we assigned
half weight).  We noticed a systematic difference between their data and
ours, so we added an offset of 4.5 \kms~ to the velocities of
\citet{blaauw1963} and 5.3 \kms~ to the velocities of \citet{abt1990} to
bring our measurements into better agreement.  Our period search revealed
two possible results, 25.4 or 27.6 d, and we investigated many possible
periods close to these values to obtain the lowest errors in the orbital
fit.  The resulting errors were two times smaller using the shorter
period, and we adopt a period of 25.4176 d.  The remaining orbital
elements for HD 1976 are listed in Table \ref{orbits}.

We expect that the speckle companion contributes about 30\% of the flux in
this system, but we saw no signs of blending in the H Balmer or
\ion{He}{1} line profiles.  However, the relatively low velocity
semiamplitude, $K$, and high $V \sin i$ of this star (see Section 4) may
prevent us from detecting the spectrum of the companion star.

\subsection{HD 14633}

In our earlier $V_r$ study of HD 14633 \citep{boyajian2005}, we determined
that this SB1 has a short period ($15.4083 \pm 0.0004$ d), high
eccentricity, and low mass function.  This system may harbor a compact
companion formed by a past supernova, making it one of the first known
quiet MXRB candidates.  In Table \ref{vel1}, we present 14 new $V_r$
measurements of this system.

To improve the orbital period, we combined a total of 103 $V_r$
measurements from \citet{bolton1978}, \citet{stone1982},
\citet{boyajian2005}, and this work.  We refine the period to 15.407 d.  
From the recent data of \citet{boyajian2005} and this work, we also
improved the orbital elements, listed in Table \ref{orbits}.  One point at
$\phi \sim 0.4$ has a very large scatter, and we assigned it zero weight in
the fit.  The new orbital elements are the same within error as our earlier
values \citep{boyajian2005}.

\subsection{HD 15137}

Likewise, HD 15137 may also be a quiet MXRB candidate \citep{boyajian2005}
which we observed to refine its orbital elements.  We measured $V_r$ for HD
15137 using the same Balmer, \ion{He}{1}, and \ion{He}{2} lines we used for
HD 14633.  Because of the broad lines in the spectrum of HD 15137, we find
significantly more scatter than in the case of HD 14633.  Our $V_r$
measurements are listed in Table \ref{vel1}.

For HD 15137, we collected 34 published $V_r$ measurements from
\citet*{conti1977} and \citet{boyajian2005} to include with our 14 new
measurements.  A period search using this complete data set suggests a
period of 30.35 d, slightly higher than our preliminary result (28.61 d;  
\citealt{boyajian2005}).  The addition of our new measurements also
affects the other orbital elements, and our revised orbital solution is
presented in Table \ref{orbits}.  However, we stress that the period of HD
15137 relies heavily on the single observation from \citet{conti1977} and
thus it remains highly uncertain.  Better orbital coverage will improve
the results significantly.

\subsection{Feige 25}

On five of the seven nights we observed Feige 25, we obtained two 30 minute
exposures which we coadded for better S/N.  On the remaining nights, we
were only able to obtain a single 30 minute exposure.  The high $\sigma$ in
our measurements likely results from the low S/N in our spectra since the
noisy \ion{He}{1} and \ion{Mg}{2} lines were difficult to fit.  While we
observe some scatter in the $V_r$ set, it is not significant and the
velocity remains constant within the measured errors.  Since our
measurements are also consistent with the values $V_r = 24.1 \pm 12.1$
\citep{martin2006} and $V_r = -5 \pm 25$ \citep{greenstein1974}, we
classify this star as radial velocity constant.

\subsection{HD 30614}

HD 30614 has been classified as an SB1 by \citet{zejnalov1986}, who found a
3.68 d period and highly eccentric ($e = 0.45$) orbit with a very low
velocity semiamplitude, $K = 9$ \kms.  Such a low $K$ would imply a very
low inclination in a massive SB1, and selection effects make it extremely
difficult to detect such systems.  We combined our 15 observations with
additional measurements of its $V_r$ from \citet{beardsley1969},
\citet{conti1977}, \citet{bohannan1978}, \citet{stone1982},
\citet{gies1986}, and \citet{zejnalov1986}.  A new period search using this
complete data set revealed the possible $P = 3.24$ d.  However, these
studies each used the mean $V_r$ from many lines, and the interline scatter
is comparable to the low $K$ found by \citet{zejnalov1986}.

Instead of using all of these data, we chose to investigate a small subset
of lines with the hope of reducing the scatter in the radial velocity
curve.  Only \citet{zejnalov1986} and Gies (private communication) provide
$V_r$ measurements for individual lines, and three lines are common to our
data sets: \ion{Si}{4} $\lambda\lambda$4089, 4116 and H$\delta$.  Since HD
30614 is a non-radial pulsator with variable stellar winds
\citep*{fullerton1996, rzayev2004}, the Balmer lines may be contaminated.  
Therefore we chose to use the mean from only the two \ion{Si}{4} lines,
which usually agree within 4 \kms~ in our data.  We repeated the
period search using this subset of data, and we found a strong signal at
$P = 3.57$ d, closer to that found by \citet{zejnalov1986}, but the low
amplitude variations probably originate in atmospheric or wind modulations
rather than orbital motion. Without the availability of higher resolution
spectra, we conclude that HD 30614 is more likely a single star that has
been misclassified.

\subsection{HD 36576}

Because the H Balmer lines have strong emission in this B2 IV-Ve star, we
used only the available \ion{He}{1} lines to measure $V_r$, listed in
Table \ref{vel2}.  Even among these lines, their asymmetric profiles
caused some difficulty in our measurements, resulting in a large scatter
for each $V_r$.  The 2AOV test confirms that the high $\sigma$ is due to
our measurement errors rather than real velocity differences between
lines.  In our sequence of observations, we find rapid yet low-amplitude
variations in $V_r$ that suggest non-radial pulsations may be present in
this star.

\subsection{HD 37737}

An orbital solution for HD 37737 was originally published by
\citet{gies1986} with an orbital period of 2.49 d.  However, our period
search using our data with theirs resulted in a very different period of
7.84 d.  Even using only Gies \& Bolton's data, we found the same period
with the modern CLEAN algorithm.  We conclude that the original period was
in error, and we present new orbital elements for this SB1 in Table
\ref{orbits}.  Figure \ref{vrmulti} shows that the \citet{gies1986} data
agree well with our data and the new orbital solution.

\subsection{HD 52266}

We measured $V_r$ for HD 52266, listed in Table \ref{vel2}, using the mean
from many H-Balmer, \ion{He}{1}, and \ion{He}{2} lines.  The star has very
broad lines that contribute to the high scatter in our measurements, and
the short term changes in $V_r$ are usually smaller than the measured
errors.  However, we find a definite increase in $V_r$ between the October
and November runs that is statistically significant in our 2AOV test.  We
suspect that HD 52266 may be a SB1 with a period too long to measure from
our data.

\subsection{HD 52533}

We initially measured $V_r$ for HD 52533 using many H Balmer, \ion{He}{1},
and \ion{He}{2} lines.  The mean velocity from the Balmer lines varies
from $3-122$ \kms, and comparable variations are seen in the \ion{He}{1}
lines. However, the \ion{He}{2} line measurements have a much higher
amplitude of variation, from $-18$ to 206 \kms!  These lines indicate
significant line-to-line variations in this star, and the 2AOV test
reveals that the probability of these variations originating from random
fluctuations within the same sample is negligible.  We suspect that this
may be a triple star system; the \ion{He}{2} line probably originates in
the hottest component, an O-type primary star in a close spectroscopic
binary.  A cooler, B-type companion is probably a widely separated,
stationary star.  Its strong \ion{He}{1} and Balmer lines cause blending
in these line profiles without affecting the \ion{He}{2} lines.  HD 52533
is not a known speckle binary.  Most recently, it was observed but not
resolved by B. Mason in 2005 November (private communication), placing an
upper limit of 30 mas on the angular separation of the wide components
assuming $\delta V < 3$.

Because of the suspected line blending, we investigate the orbit of the
primary using the mean $V_r$ from only the \ion{He}{2}
$\lambda\lambda$4200, 4542, 4686 lines.  Gies (private communication)
provided 10 measurements of these lines that we combine with our 20
observations to determine the primary's orbital period, $P = 22.19$ d.  
Only one \ion{He}{2} line was recorded in our coud\'e feed spectra, and
the large scatter in Gies's data led us to use only one or two of his
\ion{He}{2} line measurements in some cases.  For each $V_r$ based on only
one line instead of a mean, we assigned the point only half weight.  The
resulting orbit of the primary is presented in Table \ref{orbits}, and its
$V_r$ measurements are listed in Table \ref{vel1}.  We will explore this
system in more detail in a forthcoming paper.

\subsection{HD 60848}

The H$\beta$, H$\gamma$, and \ion{He}{1} $\lambda$4713 lines in HD 60848
show double peaked emission in our spectra, therefore we omitted these
lines in our $V_r$ measurements.  Like HD 36576, we found rapid, low
amplitude variations in $V_r$ in HD 60848 that are consistent with
non-radial pulsations, and we classify this star as single.  The high
scatter in our measurements is not statistically significant, and it 
is likely due to the pulsations' effect on line profiles in HD 60848.

\subsection{HD 188001}

There are a wealth of $V_r$ measurements for HD 188001 in the literature,
and several groups (\citealt{underhill1995b};  \citealt*{aslanov1984};
\citealt{aslanov1992}) have published orbital solutions for this
spectroscopic binary with periods ranging from 32.5$-$78.74 d.  Our new
observations do not agree with any of the previous solutions, so we
decided to undertake a new analysis of this star using the available data.  
Thus we gathered 14 measurements of $V_r$ from \citet{plaskett1924}, 2
from \citet{conti1977}, 11 from \citet{bohannan1978}, 8 from
\citet*{garmany1980}, 16 from \citet{stone1982}, 17 from
\citet{aslanov1984}, 20 from \citet{fullerton1990}, 9 from
\citet{underhill1995}, and 8 new values from this work (listed in Table
\ref{vel2}).

This complete sample of 105 $V_r$ values is an inhomogeneous collection of
measurements relying on different line groups and different wavelength
regions, but the measured lines are predominantly \ion{He}{1} and
\ion{He}{2} with some metal lines also included.  To retain the maximum
consistency between these data sets, whenever the authors present $V_r$
from more than one line or line group, we chose to use only \ion{He}{1}
and \ion{He}{2} lines in absorption.  These lines also provided the most
reliable $V_r$ in our own measurements; however, they are probably
affected by the stellar winds.  We observe the \ion{He}{2} $\lambda$4686
line in emission, and \citet{fullerton1990} notes a variable P Cygni
absorption trough in the \ion{He}{1} $\lambda$5876 line profiles, whereas
the \ion{C}{4} $\lambda\lambda$5801, 5812 lines in Fullerton's sample are
more likely photospheric.  Certainly we are concerned by the need to rely
on wind lines to investigate the orbital period of HD 188001, but in lieu
of a more complete data set from purely photospheric lines, we are limited
in our approach.

We performed a period search with these 105 values, eliminating spurious
frequencies that suggest periods less than 10 days since rapid $V_r$
variations are not observed in HD 188001.  The strongest signal suggests a
period $P$ = 29.83 d; however, the sinusoidal variation appears to be due
to the systematic differences between various data sets rather than true
orbital motion.  Therefore we conclude that HD 188001 is actually a single
star, possibly with variable winds that contaminate the He absorption
lines and mimic the signature of a spectroscopic binary.

\subsection{HD 195592}

We did not use the H Balmer lines to measure $V_r$ in HD 195592 due to
their inverse P Cygni profiles, suggesting that wind emission is partially
filling in these lines.  Instead we used strong, unblended \ion{He}{1},
\ion{He}{2}, and \ion{Si}{4} lines to measure velocities. We find that
$V_r$ is slightly lower during our November run than in October, and the
2AOV statistical test indicates that this difference is highly
significant.  Therefore we believe HD 195592 may be a SB1 system.  We
combined our data with 8 $V_r$ measurements from \citet{mayer1994} to
search for a possible orbital period, but the only significant signal
occured at the unphysical period $\sim 0.5$ d.  Additional data is needed
to determine if HD 195592 is indeed a binary.


\section{Spectroscopic Modeling and SED Fits}

Our O- and B-type targets include a wide range of spectral types,
luminosity types, and even emission stars, making it difficult to obtain
the temperature and surface gravity of every star.  Nonetheless, we used a
combination of sources to obtain their physical parameters.  The resulting
values, and references where applicable, are listed in Table \ref{params}.

For the B-type stars HD 1976 and Feige 25, we generated a grid of
synthetic, plane-parallel, local thermodynamic equilibrium (LTE)  
atmospheric models using the Kurucz ATLAS9 code \citep{kurucz1994}.  We
adopted solar abundances and a microturbulent velocity of 2 \kms~ for these
stars, which corresponds to the mean microturbulence observed among
late-type, main-sequence B stars \citep*{lyubimkov2004}.  Each atmospheric
model was then used to calculate a grid of model spectra using SYNSPEC
\citep{lanz2003}.  Using the measured $V_r$ from each observation, we
shifted each observed spectrum to its rest frame and created a mean
spectrum of each target to compare with the model grid.  We compared the
observed three H Balmer line profiles to model profiles convolved with a
limb-darkened, rotational broadening function and a Gaussian instrumental
broadening function to measure $T_{\rm eff}$, $\log g$, and $V \sin i$.  
For each fit, we used steps of 100 K in $T_{\rm eff}$, 0.1 dex in $\log g$,
and 10 km~s$^{-1}$ in $V \sin i$ and determined the best fit by minimizing
the square of the residual, $(O-C)^2$, over a 40 \AA~ wide region centered
on the rest wavelength of each line.  The line wings are particularly
sensitive to $\log g$ while the line depths are strong indicators of
$T_{\rm eff}$ since the strength of the Balmer lines declines steadily with
increasing $T_{\rm eff}$ in the B spectral regime.  Finally, the shape of
the line core is most sensitive to $V \sin i$, so we were able to obtain
excellent fits by adjusting these three parameters.  Using this technique,
we find errors of $\pm 100$ K in $T_{\rm eff}$ for Feige 25 and $\pm 400$ K
for HD 1976.  For both stars, the error in $\log g$ is $\pm 0.1$ and for $V
\sin i$ it is $\pm 30$ \kms.  The resulting parameters for HD 1976 and
Feige 25 are listed in columns $3-5$ of Table \ref{params}.  Our
measurements of Feige 25 are in good agreement with the physical parameters
found by \citet{keenan1983} and \citet{martin2004}.

For the hotter O-type stars, we used the OSTAR2002 grid of line-blanketed,
non-LTE, plane-parallel, hydrostatic atmosphere model spectra for O-type
stars from the TLUSTY code \citep{lanz2003}.  These models adopt solar
abundances and a fixed microturbulent velocity of 10 \kms.  Once again, we
convolved each model with a limb darkened rotational broadening function
and a Gaussian instrumental broadening function to compare with our
observed spectra.  Here we used step sizes of 100 K in $T_{\rm eff}$, 0.05 
in $\log g$, and 10 \kms~ in $V \sin i$.  
Because of the probable line blending in the Balmer and \ion{He}{1} lines 
in HD 52533 and emission partially filling the \ion{He}{1} line profiles in 
HD 60848, we used the \ion{He}{2} $\lambda\lambda$4200, 4542, 4686 lines to 
measure the physical parameters of each O-type dwarf.  The resulting 
$T_{\rm 
eff}$, $\log g$, and $V \sin i$ from the O star fits are listed in Table 
\ref{params}.  Our measurement errors are typically 400 K in $T_{\rm 
eff}$, 0.10 in $\log g$, and 10 \kms~ in $V \sin i$.

\placetable{params}

The Be star HD 36576 had obvious emission in its Balmer lines, and the O-
supergiants HD 30614, HD 188001, and HD 195592 were also difficult to fit
with the plane-parallel spectral models.  Therefore we adopted $T_{\rm
eff}$ and $\log g$ of HD 30614 and HD 36576 from \citet*{crowther2006} and
\citet{chauville2001}, respectively.  In addition, we adopted $V \sin i =
165$ \kms~ for HD 36576 \citep*{abt2002}.  We used the spectral types of
HD 188001 and HD 195592 to obtain their parameters from the calibration of 
\citet*{martins2005}.

We then compared the resulting model spectrum to the observed spectral
energy distribution (SED).  For all of our targets except Feige 25 and HD
52533, flux-calibrated UV spectra were available from the
\textit{International Ultraviolet Explorer (IUE)} archives.  We also
obtained UV fluxes in several bandpasses from \citet{thompson1978} and/or
\citet{wesselius1982} in most cases.  For HD 52266 and HD 195592, the
available \textit{IUE} fluxes disagree with the other UV fluxes, so we
prefer the UV bandpass photometry which provides a continuous SED when
combined with optical data.  Johnson $UBVR$ (from numerous sources in the
VizieR database; \citealt*{ochsenbein2000}), Str\"omgren $uvby$
\citep{hauck1998}, and 2MASS $JHK$ \citep{cutri2003} photometry were
usually available as well.  For these three filter systems, we converted
the photometric magnitudes to fluxes using the techniques of
\citet*{colina1996}, \citet{gray1998}, and \citet*{cohen2003},
respectively.  The model spectra were binned into 50 \AA~bins in order to
remove small scale line structure and compare to the observed SED.  For
those stars with close binary or speckle companions that contribute to the
SED, we neglect the companion's flux contribution in the model.

Using a grid of values for the reddening, $E(B-V)$, and the ratio of
total-to-selective extinction, $R$, we compared the reddened absolute
fluxes from the spectral models to the observed stellar flux.  We used a
step size of 0.01 for both parameters, and we determined the best fit
values of $E(B-V)$ and $R$ by minimizing $(O-C)^2$, and the typical errors
are 0.03 mag and 0.4, respectively.  The ratio of the observed stellar flux
to the reddened model flux provides the angular diameter, $\theta_{LD}$, of
each star \citep{gray1992}.  The errors in $\theta_{LD}$ are determined
from the standard deviation of the ratios from each optical and infrared
flux estimate since the ratio is wavelength independent.  These measured
parameters are listed in columns $6-8$ of Table \ref{params}.

Distances to most of our targets were obtained from
\citet{vansteenberg1988}, \citet{mason1998}, \citet*{patriarchi2003}, and
\citet{martin2006}.  For the remaining stars HD 1976 and HD 36576, we
obtained an estimate of their absolute magnitudes from \citet{wegner2000}
and combined these with the observed $V$ and our calculated $A(V) = R
\times E(B-V)$ to determine their distances.  Finally, we used
$\theta_{LD}$ with the known distances to determine the radius, $R_\star$,
for each star (column 9 of Table \ref{params}).  We note that the 
closest stars in our list, HD 1976 and HD 36576, also have 
\textit{Hipparcos} parallaxes available, but the larger distances from 
these measurements (420 and 570 pc, respectively) imply significantly 
larger $R_\star$ than we expect for these stars.

In Table \ref{params} we also provide the predicted $R_\star$ from the
calibration of \citet{martins2005} for stars with $T_{\rm eff} > 30000$,
and for cooler stars we use the evolutionary models of \citet{schaller1992}
to determine the expected value.  The observed differences between the
derived and predicted radii are usually small, and we find no systematic
differences between the binaries' and single stars' radii that may be due
to neglecting the companion fluxes in the SED fits.  Note that we do find
large discrepancies in $R_\star$ for the binaries HD 37737 and HD 195592,
but we find a similar error for the single star HD 188001.  All three stars
are giants and supergiants, and the errors in their predicted physical
parameters are larger \citep{martins2005}.  In these cases the distances
require a significant downward revision to match our derived radii with the
expected values.  Therefore we derive a new distance to each target based
upon the expected $R_\star$ and our measured $\theta_{\rm LD}$, and we
adopt the mean of these two distances with errors representing the range of
possible values.  The final adopted distances are listed in Table
\ref{params}.

We also inspected the observed SED to look for an IR excess in our
targets, which may be indicative of an accretion disk around a compact
companion.  The emission-line stars HD 36576 and HD 60848 appear to have a
small $K$-band flux excess due to their circumstellar disks, but no other
targets show evidence of an IR excess.


\section{Runaway Status of Targets}

Our new radial velocity data and the distance constraints above allow us
to reexamine the runaway status of each of our targets.  We list in Table
\ref{spacevel}, columns $2-3$ the galactic longitude, $\ell$, and
latitude, $b$, for each star.  Columns $4-5$ give the proper motions,
$\mu_\alpha \cos \delta$ and $\mu_\delta$, obtained from The Second U.S. Naval
Observatory CCD Astrograph Catalog (UCAC2; \citealt{zacharias2004}) and
The UCAC2 Bright Star Supplement \citep*{urban2004}.  We computed the
peculiar transverse and radial space velocity, $V_{t,pec}$ and
$V_{r,pec}$, of each star using the method of \citet{berger2001}, and
these are listed in columns $6-7$.  These are combined into a single
measure of the peculiar space velocity, $V_{pec}$, in column 8.  Our
values of $V_{t,pec}$ and $V_{r,pec}$ for HD 37737 and HD 52533 agree 
closely with those found by \citet{moffat1998}.

We classify as definite runaways those stars with $V_{pec} > 30$ \kms~plus
the error in $V_{pec}$ \citep{gies1986}, and we find seven such stars among
our sample: HD 14633, HD 15137, Feige 25, HD 30614, HD 36576, HD 188001,
and HD 195592.  Our findings agree with those of \citet*{noriega1997}, who
found bow shocks associated with HD 30614, HD 188001, and HD 195592 due to
their fast space velocities.  Uncertainties in the proper motions and
distances contribute to large errors in $V_{pec}$ for several of our stars,
so the tentative runaways HD 37737 and HD 52533 also have $V_{pec} > 30$
\kms~ within their errors.  Having a lower $V_{pec}$ does not exclude the
possiblity that a star was ejected in the past -- as long as its velocity
is greater than the escape velocity from the cluster, an ejected star will
effectively be a runaway.

\placetable{spacevel}

Although we find that HD 37737 is tentatively a runaway, it is worthwhile
to note that the star has also been classified as a member of the Aur OB1
association \citep{humphreys1978}.  It may also be surrounded by a
symmetric \ion{H}{2} bubble \citep{noriega1997}, although the morphology of
the bubble is difficult to classify with certainty. To investigate whether
HD 37737 is a member of Aur OB1, we compared its proper motion to those of
the other proposed Aur OB1 members using the UCAC2 and UCAC2 Bright Star
Supplement catalogs.  We find that HD 37737 and 7 other stars from
Humphrey's list do indeed have similar proper motions, although the
remaining 7 stars are moving in nearly an opposite direction.  From these
results we doubt that the Aur OB1 association is truely a bound group, but
the evidence does suggest that multiple open clusters may be superimposed
upon this region of the sky.  Further investigation of the groups'
distances are required to determine whether HD 37737 is a member.

The stars HD 1976, HD 52266, and HD 60848 have low $V_{pec}$ and are
probably not runaway stars.  HD 52266 (along with the higher velocity
systems HD 52533 and HD 195592; \citealt{dewit2004}) may be a member of
a previously undetected cluster.  Since our spectra of HD 52533 suggest
that it may be a triple system, it is very likely that a sparse cluster
indeed exists and that this system is not a runaway.  We will explore this
possibility in the future.

Our search was designed to detect runaway binary systems, and it is no
surprise that we identify several in our sample (HD 14633, HD 15137, HD
195592, and the possible runaway HD 37737).  Both dynamical interactions
within a dense open cluster and a supernova in a close binary can produce
runaway SB1 systems, and our results highlight how rare such systems are.  
These ejection mechanisms can be distinguished observationally, and we
explore both scenarios for HD 14633 and HD 15137 in a forthcoming paper.


\acknowledgments

We thank the referee for his/her careful consideration of this work.  We 
are grateful to Di Harmer for her assistance at the telescope, and we also
thank John Martin and Alex Fullerton for providing their thesis data and
for their helpful comments on this work.  Kathy Vieira, Bill van Altena,
and Norbert Zacharias contributed useful advice about the proper motion
analysis.  This material is based on work supported by the National
Science Foundation under Grants No.~AST-0205297, AST-0401460, and
AST-0506573.  Institutional support has been provided from the GSU College
of Arts and Sciences and from the Research Program Enhancement fund of the
Board of Regents of the University System of Georgia, administered through
the GSU Office of the Vice President for Research.  Some of the data
presented in this paper were obtained from the Multimission Archive at the
Space Telescope Science Institute (MAST).  STScI is operated by the
Association of Universities for Research in Astronomy, Inc., under NASA
contract NAS5-26555. Support for MAST for non-HST data is provided by the
NASA Office of Space Science via grant NAG5-7584 and by other grants and
contracts.

Facilities: \facility{KPNO, IUE(SWP, LWP, LWR)}


\clearpage



\clearpage
\begin{deluxetable}{lcl}
\rotate
\tabletypesize{\scriptsize}
\tablewidth{0pt}
\tablecaption{Lines Used for Radial Velocities\label{lines}}
\tablehead{
\colhead{Star} &
\colhead{Telescope} &
\colhead{Lines} } 
\startdata
HD 1976    & 2.1~m    &  H$\beta$, H$\gamma$, H$\delta$, 
		\ion{He}{1} $\lambda\lambda$4144, 4388, 4471, 4713, 4922,
		\ion{C}{2} $\lambda$4267,
		\ion{Mg}{2} $\lambda$4481 \\

           & coud\'e &  H$\gamma$, \ion{He}{1} $\lambda\lambda$4388, 4471 \\
\\
HD 14633   & 2.1~m    &  H$\beta$, H$\gamma$, H$\delta$;
		\ion{He}{1} $\lambda\lambda$4121, 4144, 4388, 4471, 
		4713 (Oct. only), 4922; 
		\ion{He}{2} $\lambda\lambda$4200, 4542, 4686; 
		\ion{Si}{4} $\lambda$4116; 
		\ion{N}{3} $\lambda$4640\tablenotemark{a} \\
\\
HD 15137   & 2.1~m    &  H$\beta$, H$\gamma$, H$\delta$, 
		\ion{He}{1} $\lambda\lambda$4144, 4388, 4471, 4713, 4922;
		\ion{He}{2} $\lambda\lambda$4200, 4542, 4686;
		\ion{Si}{4} $\lambda$4089;
		\ion{N}{3} $\lambda$4379 \\
\\
Feige 25   & 2.1~m    &  H$\beta$, H$\gamma$, H$\delta$; 
		\ion{He}{1} $\lambda\lambda$4144, 4388, 4471, 4922 (Oct. only); 
		\ion{Mg}{2} $\lambda$4481 \\

           & coud\'e &  H$\gamma$; \ion{He}{1} $\lambda\lambda$4388, 
		4471  \\
\\
HD 30614   & 2.1~m    &  H$\beta$, H$\gamma$, H$\delta$;
		\ion{He}{1} $\lambda\lambda$4121, 4144, 4388, 4471, 4713, 
		4921; 
		\ion{He}{2} $\lambda\lambda$4200, 4542; 
		\ion{Si}{4} $\lambda\lambda$4089, 4116;
		\ion{N}{3} $\lambda$4379 \\

           & coud\'e &  H$\delta$, \ion{He}{1} $\lambda\lambda$4388, 
		4471; \ion{He}{2} $\lambda$4542; \ion{N}{3} $\lambda$4379 \\
\\
HD 36576   & 2.1~m    &  \ion{He}{1} $\lambda\lambda$4121, 4144, 4388, 
		4471, 4713 \\

           & coud\'e &  \ion{He}{1} $\lambda\lambda$4388, 4471  \\
\\
HD 37737   & 2.1~m    &  H$\beta$, $H\gamma$, H$\delta$; 
		\ion{He}{1} $\lambda\lambda$4144, 4388, 4471, 4713, 4922; 
		\ion{He}{2} $\lambda\lambda$4200, 4542, 4686; 
		\ion{Si}{4} $\lambda$4089 \\

           & coud\'e &  $H\gamma$; \ion{He}{1} $\lambda\lambda$4388, 4471; 
		\ion{He}{2} $\lambda$ 4542 \\
\\
HD 52266   & 2.1~m    &  H$\beta$, $H\gamma$, H$\delta$;
		\ion{He}{1} $\lambda\lambda$4144, 4388, 4471, 4713, 4921;
		\ion{He}{2} $\lambda\lambda$4200, 4542, 4686;
		\ion{Si}{4} $\lambda$4089 \\

           & coud\'e &  $H\gamma$; \ion{He}{1} $\lambda\lambda$4388, 
		4471 \\
\\
HD 52533   & 2.1~m    &  \ion{He}{2} $\lambda\lambda$4200, 4542, 4686 \\

           & coud\'e &  \ion{He}{2} $\lambda$4542 \\
\\
HD 60848   & 2.1~m    &  H$\delta$; 
		\ion{He}{1} $\lambda\lambda$4144, 4388, 4471, 4921;
		\ion{He}{2} $\lambda\lambda$4200, 4542, 4686; 
		\ion{Si}{4} $\lambda$4089;
		\ion{N}{3} $\lambda$4379 \\

           & coud\'e &  \ion{He}{1} $\lambda\lambda$4388, 4471; 
		\ion{He}{2} $\lambda$4542 \\
\\
HD 188001  & 2.1~m    &  \ion{He}{1} $\lambda\lambda$ 4120, 4144, 4388, 
		4471, 4713, 4922; 
		\ion{He}{2} $\lambda\lambda$4200, 4542 \\

           & coud\'e &  \ion{He}{1} $\lambda\lambda$4388, 4471; 
		\ion{He}{2} $\lambda$4542 \\
\\
HD 195592  & 2.1~m    &  \ion{He}{1} $\lambda\lambda$4120, 4143, 4387, 
		4471, 4713, 4922; 
		\ion{He}{2} $\lambda\lambda$4200, 4542; 
		\ion{Si}{4} $\lambda\lambda$4089, 4116; 
		\ion{N}{3} $\lambda\lambda$4379, 4640\tablenotemark{a} \\

           & coud\'e &  \ion{He}{1} $\lambda\lambda$4388, 4471; 
		\ion{He}{2} $\lambda$ 4542 \\
\enddata
\tablenotetext{a}{Blend.}
\end{deluxetable}

\clearpage
\begin{deluxetable}{lcccccc}
\tablewidth{0pt}
\tablecaption{Results from Two-way Analysis of Variance Test\label{2aov}}
\tablehead{
\colhead{ } &
\colhead{Degrees of} &
\colhead{ } &
\colhead{ } &
\colhead{Degrees of} &
\colhead{ } &
\colhead{ } \\
\colhead{Star} &
\colhead{Freedom (LL)} &
\colhead{$F$(LL)} &
\colhead{$p$(LL) (\%)} &
\colhead{Freedom (NN)} &
\colhead{$F$(NN)} &
\colhead{$p$(NN) (\%)} 
}
\startdata
HD 1976    &     10 & \phn 2.54  &  \phn 0.78  &     13 & \phn    13.45  &  0.00  \\
HD 14633   &     13 & \phn 1.96  &  \phn 2.72  &     13 & \phn    33.65  &  0.00  \\
HD 15137   &     12 & \phn 5.17  &  \phn 0.00  &     13 & \phn\phn 3.18  &  0.03  \\
Feige 25   & \phn 7 & \phn 1.43  &      23.32  & \phn 4 & \phn\phn 2.74  &  4.84  \\
HD 30614   &     13 &     11.07  &  \phn 0.00  &     13 & \phn    12.17  &  0.00  \\
HD 36576   & \phn 4 & \phn 2.08  &  \phn 9.87  &     12 & \phn\phn 8.77  &  0.00  \\
HD 37737   &     11 &     14.39  &  \phn 0.00  &     12 &        316.90  &  0.00  \\
HD 52266   &     11 & \phn 1.84  &  \phn 5.22  &     14 & \phn    10.38  &  0.00  \\
HD 52533   & \phn 9 & \phn 6.48  &  \phn 0.00  &     14 & \phn    13.30  &  0.00  \\
HD 60848   & \phn 9 &     11.35  &  \phn 0.00  &     13 & \phn\phn 2.22  &  1.27  \\
HD 188001  & \phn 7 & \phn 1.54  &      18.09  & \phn 6 & \phn    11.36  &  0.00  \\
HD 195592  &     11 & \phn 2.05  &  \phn 3.41  & \phn 7 & \phn\phn 7.27  &  0.00  \\
\enddata
\end{deluxetable}

\clearpage
\begin{deluxetable}{lccccccccc}
\rotate
\tabletypesize{\scriptsize}
\tablewidth{0pt}
\tablecaption{Orbital Elements\label{orbits}}
\tablehead{
\colhead{ } &
\colhead{$P$} &
\colhead{$T$} &
\colhead{ } &
\colhead{$\omega$} &
\colhead{$K_1$} &
\colhead{$\gamma$} &
\colhead{$f(m)$} &
\colhead{$a_1 \sin i$} &
\colhead{$\sigma$} \\
\colhead{Star} &
\colhead{(days)} &
\colhead{(HJD--2,400,000)} &
\colhead{$e$} &
\colhead{(deg)} &
\colhead{(km s$^{-1}$)} &
\colhead{(km s$^{-1}$)} &
\colhead{($M_\odot$)} &
\colhead{($R_\odot$)} &
\colhead{(km s$^{-1}$)} }
\startdata
HD 1976    &  $25.4176 \pm 0.0004$          &  $35783.5 \pm 0.1$      &  $0.12 \pm 0.03$  &  $172 \pm 2$  &
$\phn23.6 \pm 0.7$  &  \phn $-9.7 \pm 0.5$  &  $0.034 \pm 0.003$  &     $11.7 \pm 0.4$  &  $\phn4.81$  \\

HD 14633   &  $15.4082 \pm 0.0004$  &  $51885.07 \pm 0.01$  &  $0.700 \pm 0.003$  &  $138.7 \pm 0.5$     &  
$\phn19.0 \pm 0.1$  &  $-38.17 \pm 0.08$  &  $0.0040 \pm 0.0001$  &  \phn$4.13 \pm 0.04$  &  $\phn1.96$  \\ 

HD 15137   &  $30.35$\tablenotemark{a}      &  $51879.7 \pm 0.6$    &  $0.48 \pm 0.07$  &  $\phn148 \pm 10$  &
$\phn13 \pm 1$  &  $-48.4 \pm 0.6$  &  $0.004 \pm 0.002$  &  \phn$6.7 \pm 0.8$  &  $\phn3.94$  \\

HD 37737   & \phn $7.8400 \pm 0.0002$     &  $53690.2 \pm 0.1$    &  $0.43 \pm 0.02$  &  $158 \pm 5$  &  
$\phn72 \pm 2$  &  $\phn-8 \pm 2\phn$  &  $0.22 \pm 0.02$  &  $10.1 \pm 0.3$  &  $\phn5.28$ \\ 

HD 52533   &  $22.1861 \pm 0.0002$        &  $44635.01 \pm 0.03$  &  $0.30 \pm 0.01$  &  $330.3 \pm 0.5$  &  
$   105 \pm 1$  &  \phs$76.7 \pm 0.6$  &  $2.30 \pm 0.07$  &  $43.8 \pm 0.4$  &  $11.7\phn$  \\
\enddata
\tablenotetext{a}{Fixed.}
\end{deluxetable}

\clearpage
\begin{deluxetable}{lcccclcccc}
\tabletypesize{\scriptsize}
\tablewidth{0pt}
\tablecaption{Radial Velocity Measurements of Binaries with 
Orbital Solutions\label{vel1}}
\tablehead{
\colhead{ } & 
\colhead{HJD} &
\colhead{Orbital}       &
\colhead{$V_r$} &
\colhead{$(O-C)$} &
\colhead{ } &
\colhead{HJD} &
\colhead{Orbital}       &
\colhead{$V_r$} &
\colhead{$(O-C)$}       \\
\colhead{Star} &
\colhead{($-$2,450,000)} &
\colhead{Phase}         &
\colhead{(km s$^{-1}$)} &
\colhead{(km s$^{-1}$)} &
\colhead{Star} &
\colhead{($-$2,450,000)} &
\colhead{Phase}         &
\colhead{(km s$^{-1}$)} &
\colhead{(km s$^{-1}$)} }
\startdata
HD 1976     &  3657.787  &  0.225 & \phn         $ -14.2$ &\phn     $  -0.5$  &  HD 15137  &  3696.628  &  0.864 & \phn         $ -48.8$ &\phn     $  -4.4$  \\
            &  3658.833  &  0.266 & \phn\phn     $  -4.8$ &\phn\phs $   3.0$  &            &  3696.841  &  0.871 & \phn         $ -38.3$ &\phn\phs $   6.5$  \\
            &  3658.835  &  0.267 & \phn\phn     $  -4.3$ &\phn\phs $   3.5$  &            &             &        &                       &                   \\
            &  3659.720  &  0.301 & \phn\phn     $  -0.3$ &\phn\phs $   3.0$  &  HD 37737  &  3658.005  &  0.898 & \phn         $ -13.1$ &\phn\phs $   5.2$ \\
            &  3659.721  &  0.301 & \phn\phn     $  -0.1$ &\phn\phs $   3.2$  &            &  3658.983  &  0.023 &              $-108.8$ &\phn     $  -0.2$ \\
            &  3660.754  &  0.342 & \phn\phn\phs $   3.0$ &\phn\phs $   1.7$  &            &  3659.916  &  0.142 & \phn         $ -50.9$ &\phn\phs $   0.2$ \\
            &  3663.735  &  0.459 & \phn\phs     $  13.7$ &\phn\phs $   4.0$  &            &  3662.988  &  0.533 & \phn\phs     $  31.4$ &\phn     $  -0.7$ \\
 &  3685.735\tablenotemark{a}  &  0.325 & \phn\phn   $ -9.4$ &\phn  $  -8.9$  &            &  3663.016  &  0.537 & \phn\phs     $  27.4$ &\phn     $  -4.9$ \\
            &  3693.635  &  0.636 & \phn\phs     $  13.8$ &\phn\phs $   6.3$  &            &  3663.980  &  0.660 & \phn\phs     $  30.1$ &\phn     $  -5.0$ \\
            &  3693.790  &  0.642 & \phn\phn\phs $   7.3$ &\phn\phs $   0.2$  &            &  3686.830\tablenotemark{a}  &  0.575 & \phn\phs     $  27.2$ &\phn     $  -6.7$ \\
            &  3694.701  &  0.678 & \phn\phn\phs $   6.9$ &\phn\phs $   2.6$  &            &  3687.030\tablenotemark{a}  &  0.600 & \phn\phs     $  34.7$ &\phn     $  -0.1$ \\
            &  3695.674  &  0.716 & \phn\phn     $  -1.4$ &\phn     $  -1.8$  &            &  3687.832\tablenotemark{a}  &  0.702 & \phn\phs     $  31.6$ &\phn     $  -2.0$ \\
            &  3695.769  &  0.720 & \phn\phn     $  -0.8$ &\phn     $  -0.7$  &            &  3688.037\tablenotemark{a}  &  0.729 & \phn\phs     $  30.7$ &\phn     $  -1.1$ \\
            &  3696.565  &  0.751 & \phn\phn     $  -4.9$ &\phn     $  -0.9$  &            &  3688.900\tablenotemark{a}  &  0.839 & \phn\phs     $  14.5$ &\phn\phs $   3.8$ \\
            &  3696.789  &  0.760 & \phn\phn     $  -2.0$ &\phn\phs $   3.2$  &            &  3689.028\tablenotemark{a}  &  0.855 & \phn\phn\phs $   2.6$ &\phn     $  -2.0$ \\
            &            &        &                       &                   &            &  3693.864  &  0.472 & \phn\phs     $  23.7$ &\phn     $  -3.7$ \\
HD 14633    &  3657.799  &  0.051 & \phn         $ -60.0$ &\phn     $  -0.2$  &            &  3693.988  &  0.488 & \phn\phs     $  37.4$ &\phn\phs $   8.6$ \\
            &  3657.803  &  0.051 & \phn         $ -59.4$ &\phn\phs $   0.3$  &            &  3694.886  &  0.602 & \phn\phs     $  44.8$ &\phn\phs $   9.9$ \\
            &  3658.843  &  0.118 & \phn         $ -50.1$ &\phn     $  -1.1$  &            &  3695.868  &  0.727 & \phn\phs     $  33.4$ &\phn\phs $   1.5$ \\
            &  3659.756  &  0.178 & \phn         $ -40.0$ &\phn\phs $   4.3$  &            &  3695.985  &  0.742 & \phn\phs     $  36.0$ &\phn\phs $   5.5$ \\
            &  3660.760  &  0.243 & \phn         $ -47.5$ &\phn     $  -6.6$  &            &  3696.737  &  0.838 & \phn\phs     $  11.2$ &\phn\phs $   0.3$ \\
            &  3663.739  &  0.436 & \phn         $ -32.8$ &\phn\phs $   2.4$  &            &  3697.012  &  0.873 & \phn         $ -12.5$ &\phn     $  -8.6$ \\
            &  3693.639  &  0.377 & \phn         $ -36.7$ &\phn     $  -0.1$  &            &             &        &                       &                  \\
            &  3693.852\tablenotemark{b}  & 0.391 & \phn $ -49.0$ & $ -12.7$  &  HD 52533  &  3658.024  &  0.697 & \phn         $ -17.9$ &         $ -19.8$ \\
            &  3694.705  &  0.446 & \phn         $ -33.4$ &\phn\phs $   1.6$  &            &  3659.020  &  0.742 & \phn\phs     $  21.2$ &\phs     $  11.8$ \\
            &  3694.863  &  0.456 & \phn         $ -36.1$ &\phn     $  -1.3$  &            &  3659.970  &  0.785 & \phn\phs     $  30.4$ &\phn\phs $   8.1$ \\
            &  3695.679  &  0.509 & \phn         $ -37.1$ &\phn     $  -3.3$  &            &  3659.978  &  0.786 & \phn\phs     $  33.2$ &\phs     $  10.8$ \\
            &  3695.875  &  0.522 & \phn         $ -32.3$ &\phn\phs $   1.2$  &            &  3662.994  &  0.921 & \phs         $ 119.4$ &\phn     $  -0.6$ \\
            &  3696.624  &  0.570 & \phn         $ -36.8$ &\phn     $  -4.2$  &            &  3663.020  &  0.923 & \phs         $ 118.7$ &\phn     $  -2.5$ \\
            &  3696.838  &  0.584 & \phn         $ -29.6$ &\phn\phs $   2.7$  &            &  3663.994  &  0.967 & \phs         $ 160.6$ &\phn     $  -6.4$ \\
            &            &        &                       &                   &            &  3684.995\tablenotemark{a,c}  &  0.913 & \phs         $ 108.9$ &\phn     $  -2.5$ \\
HD 15137    &  3657.812  &  0.586 & \phn         $ -40.8$ &\phn\phs $   0.8$  &            &  3686.885\tablenotemark{a,c}  &  0.998 & \phs         $ 193.8$ &\phn\phs $   0.6$ \\
            &  3658.850  &  0.620 & \phn         $ -45.7$ &\phn     $  -4.5$  &            &  3687.892\tablenotemark{a,c}  &  0.044 & \phs         $ 205.8$ &\phn     $  -2.3$ \\
            &  3659.806  &  0.651 & \phn         $ -41.2$ &\phn     $  -0.2$  &            &  3688.877\tablenotemark{a,c}  &  0.088 & \phs         $ 169.8$ &         $ -27.0$ \\
            &  3659.810  &  0.651 & \phn         $ -36.6$ &\phn\phs $   4.4$  &            &  3689.006\tablenotemark{a,c}  &  0.094 & \phs         $ 181.2$ &         $ -12.8$ \\
            &  3660.765  &  0.683 & \phn         $ -47.0$ &\phn     $  -6.1$  &            &  3693.875  &  0.313 & \phn\phs     $  67.4$ &\phn     $  -1.3$ \\
            &  3663.742  &  0.781 & \phn         $ -40.9$ &\phn\phs $   0.5$  &            &  3693.998  &  0.319 & \phn\phs     $  62.8$ &\phn     $  -3.6$ \\
            &  3693.644  &  0.766 & \phn         $ -39.9$ &\phn\phs $   1.3$  &            &  3694.894  &  0.359 & \phn\phs     $  46.9$ &\phn     $  -3.7$ \\
            &  3693.855  &  0.773 & \phn         $ -47.7$ &\phn     $  -6.4$  &            &  3695.023  &  0.365 & \phn\phs     $  47.8$ &\phn     $  -0.8$ \\
            &  3694.713  &  0.801 & \phn         $ -38.6$ &\phn\phs $   3.2$  &            &  3695.896  &  0.404 & \phn\phs     $  30.4$ &\phn     $  -5.4$ \\
            &  3694.876  &  0.807 & \phn         $ -42.0$ &\phn     $  -0.0$  &            &  3696.002  &  0.409 & \phn\phs     $  37.6$ &\phn\phs $   3.3$ \\
            &  3695.683  &  0.833 & \phn         $ -47.9$ &\phn     $  -5.1$  &            &  3696.825  &  0.446 & \phn\phs     $  43.2$ &\phs     $  18.9$ \\
            &  3695.887  &  0.840 & \phn         $ -37.9$ &\phn\phs $   5.2$  &            &  3697.019  &  0.455 & \phn\phs     $  28.1$ &\phn\phs $   5.9$ \\
\enddata
\tablenotetext{a}{Obtained with the coud\'e feed telescope}
\tablenotetext{b}{Measurement assigned zero weight}
\tablenotetext{c}{Measurement assigned half weight}
\end{deluxetable}

\clearpage
\begin{deluxetable}{lccclccc}
\tabletypesize{\scriptsize}
\tablewidth{0pt}
\tablecaption{Radial Velocity Measurements of Other Stars\label{vel2}}
\tablehead{
\colhead{ } &
\colhead{HJD} &
\colhead{$V_r$} & 
\colhead{$\sigma$} &
\colhead{ } &
\colhead{HJD} &
\colhead{$V_r$} &
\colhead{$\sigma$}  \\
\colhead{Star} &
\colhead{($-$2,450,000)} &
\colhead{(km s$^{-1}$)} &
\colhead{(km s$^{-1}$)} &
\colhead{Star} &
\colhead{($-$2,450,000)} &
\colhead{(km s$^{-1}$)} &
\colhead{(km s$^{-1}$)}
}
\startdata
Feige 25           &  3657.834\tablenotemark{a}  &  \phs     34.3  &      10.4  &  HD 52266           &  3663.991                   &  \phs     20.5  &     10.9  \\
                   &  3658.867\tablenotemark{a}  &  \phs     36.0  & \phn  5.6  &                     &  3684.985\tablenotemark{c}  &  \phs     30.2  &     18.4  \\
                   &  3663.749                   &  \phs     47.3  &      11.4  &                     &  3693.871                   &  \phs     39.0  & \phn 8.2  \\
                   &  3688.815\tablenotemark{b}  &  \phs     25.3  &      16.3  &                     &  3693.993                   &  \phs     39.2  &     12.6  \\
                   &  3693.838                   &  \phs     39.4  & \phn  7.1  &                     &  3694.891                   &  \phs     33.7  & \phn 6.4  \\
                   &  3696.689\tablenotemark{a}  &  \phs     43.3  &      12.9  &                     &  3695.019                   &  \phs     33.5  & \phn 5.8  \\
                   &                             &                 &            &                     &  3695.893                   &  \phs     32.7  & \phn 9.1  \\
HD 30614           &  3657.989                   &  \phs\phn  8.9  &      12.5  &                     &  3695.999                   &  \phs     29.5  &     13.9  \\
                   &  3658.906                   &  \phs\phn  9.0  & \phn  6.3  &                     &  3696.822                   &  \phs     31.7  &     13.4  \\
                   &  3659.010                   &  \phs\phn  3.2  & \phn  6.8  &                     &  3697.016                   &  \phs     38.8  & \phn 9.4  \\
                   &  3659.899                   &  \phs     22.7  &      10.2  &                     &                             &                 &           \\
                   &  3659.901                   &  \phs     10.5  & \phn  8.6  &  HD 60848           &  3658.032                   &  \phs     13.5  &     14.1  \\
                   &  3662.980                   &  \phs     22.5  & \phn  7.9  &                     &  3659.026                   &  \phs     20.7  &     14.9  \\
                   &  3663.976                   &  \phs     13.3  &      16.2  &                     &  3659.987                   &  \phs     18.9  &     17.8  \\
                   &  3686.859\tablenotemark{c}  &  \phs\phn  2.2  &      11.4  &                     &  3659.995                   &  \phs     26.6  &     15.0  \\
                   &  3693.858                   &  \phs\phn  2.9  & \phn  7.4  &                     &  3663.002                   &  \phs     19.7  &     13.6  \\
                   &  3693.983                   &  \phs\phn  1.6  & \phn  9.7  &                     &  3663.999                   &  \phs     21.4  &     14.3  \\
                   &  3694.879                   &  \phs     22.7  & \phn  9.1  &                     &  3685.004\tablenotemark{c}  &  \phs     33.8  &     15.4  \\
                   &  3695.000                   &  \phs     13.3  & \phn  9.3  &                     &  3693.879                   &  \phs     10.2  &     12.7  \\
                   &  3695.861                   &  \phs     18.4  &      10.7  &                     &  3694.001                   &  \phs     26.0  &     16.7  \\
                   &  3695.978                   &  \phs     15.4  &      14.6  &                     &  3694.897                   &  \phs     26.8  &     14.4  \\
                   &  3696.731                   &  \phs\phn  9.8  & \phn  8.6  &                     &  3695.026                   &  \phs     18.5  &     13.4  \\
                   &                             &                 &            &                     &  3695.898                   &  \phs     21.2  & \phn 7.4  \\
HD 36576           &  3657.996                   &  \phs     62.7  &      28.8  &                     &  3696.005                   &  \phs     26.1  & \phn 9.7  \\
                   &  3658.978                   &  \phs     37.8  & \phn  6.7  &                     &  3696.827                   &  \phs     21.9  &     14.5  \\
                   &  3659.911                   &  \phs     21.4  & \phn  7.2  &                     &  3697.021                   &  \phs     17.5  &     10.2  \\
                   &  3662.983                   &  \phs     44.6  & \phn  7.7  &                     &                             &                 &           \\
                   &  3663.985                   &  \phs     35.1  & \phn  8.2  &  HD 188001          &  3657.649                   &  \phs     16.9  &  \phn  6.3  \\
                   &  3686.866\tablenotemark{c}  &  \phs     62.1  &      23.6  &                     &  3661.609                   &  \phs     20.8  &  \phn  7.6  \\
                   &  3687.913\tablenotemark{c}  &  \phs     74.2  &      39.0  &                     &  3663.570                   &  \phs     21.8  &  \phn  8.5  \\
                   &  3693.869                   &  \phs     23.5  &      11.4  &                     &  3687.603\tablenotemark{c}  &  \phs     11.8  &  \phn  9.4  \\
                   &  3693.992                   &  \phs     27.4  &      10.1  &                     &  3693.551                   &  \phs\phn  6.3  &  \phn  5.3  \\
                   &  3694.890                   &  \phs     45.1  & \phn  9.5  &                     &  3694.561                   &  \phs\phn  6.2  &  \phn  6.5  \\
                   &  3695.019                   &  \phs     55.4  & \phn  8.7  &                     &  3695.556                   &  \phs     16.0  &  \phn  4.3  \\
                   &  3695.871                   &  \phs     58.3  & \phn  6.7  &                     &  3696.553                   &  \phs     25.3  &  \phn  6.1  \\
                   &  3695.988                   &  \phs     53.9  & \phn  5.4  &                     &                             &                 &           \\
                   &  3696.741                   &  \phs     61.4  & \phn  7.2  &  HD 195592          &  3657.668                   &        $-24.5$  &      12.2 \\
                   &  3697.009                   &  \phs     34.8  & \phn  8.9  &                     &  3657.671                   &        $-23.2$  &  \phn 3.9 \\
                   &                             &                 &            &                     &  3661.615                   &        $-19.3$  &  \phn 7.4 \\
HD 52266           &  3658.013                   &  \phs     14.9  &      18.5  &                     &  3663.574                   &        $-19.0$  &  \phn 7.8 \\
                   &  3659.013                   &  \phs     17.2  & \phn  6.7  &                     &  3688.620\tablenotemark{c}  &        $-20.4$  &  \phn 3.2 \\
                   &  3659.956                   &  \phs     15.4  & \phn  6.7  &                     &  3693.556                   &        $-28.7$  &  \phn 6.8 \\
                   &  3659.961                   &  \phs     16.5  & \phn  6.8  &                     &  3694.566                   &        $-26.8$  &  \phn 5.3 \\
                   &  3662.998                   &  \phs     24.2  & \phn  7.9  &                     &  3695.561                   &        $-32.1$  &  \phn 4.8 \\
                   &  3663.026                   &  \phs     12.7  &      15.1  &                     &  3696.558                   &        $-32.8$  &  \phn 6.2 \\
\enddata
\tablenotetext{a}{Obtained using co-added spectra from 2.1~m telescope}
\tablenotetext{b}{Obtained using co-added spectra from coud\'e feed telescope}
\tablenotetext{c}{Obtained with the coud\'e feed telescope}
\end{deluxetable}

\clearpage
\begin{deluxetable}{llclccclccclc}
\rotate
\tabletypesize{\scriptsize}
\tablewidth{0pt}
\tablecaption{Stellar Physical Parameters\label{params}}
\tablehead{
\colhead{Star} &
\colhead{Spectral} &
\colhead{$T_{\rm eff}$} &
\colhead{ } &
\colhead{$V \sin i$} &
\colhead{ } &
\colhead{ } &
\colhead{$\theta_{\rm LD}$} &
\colhead{$d$} &
\colhead{derived} &
\colhead{expected} &
\colhead{adopted} &
\colhead{ } \\
\colhead{Star} &
\colhead{Type} &
\colhead{(K)} &
\colhead{$\log g$} &
\colhead{(\kms)} &
\colhead{$E(B-V)$} &
\colhead{$R$} &
\colhead{($\mu$as)} &
\colhead{(pc)} &
\colhead{$R_{\star}$ ($R_\odot$)} &
\colhead{$R_{\star}$ ($R_\odot$)} &
\colhead{$d$ (pc)} &
\colhead{References} }
\startdata
HD 1976    &  B5 IV       &  16100  &  3.8 \phn  &     160  &  $0.13 \pm 0.02$  &  $2.60 \pm 0.5$  &     184 $\pm$  9 \phn & 
    \phn 190  & \phn      3.8 $\pm$ 0.2  & \phn 5.7  & \phn 240 $\pm$ 50  \phn &  9        \\
HD 14633   &  O8.5 V      &  35100  &  3.95  &     138  &  $0.13 \pm 0.02$  &  $3.18 \pm 0.5$  & \phn 40 $\pm$  2 \phn & 
        2150  & \phn      9.2 $\pm$ 0.4  & \phn 8.3  &     2040 $\pm$ 110      &  6, 10     \\
HD 15137   &  O9.5 V      &  29700  &  3.50  &     234  &  $0.43 \pm 0.04$  &  $3.18 \pm 0.5$  & \phn 56 $\pm$  7 \phn &
        2650  &          16.1 $\pm$ 1.9  &     13.2  &     2420 $\pm$ 230      &  6, 10     \\
Feige 25   &  B7          &  13200  &  4.0 \phn  &     250  &  $0.12 \pm 0.01$  &  $2.94 \pm 0.1$  & \phn 11 $\pm$ 0.1     &
        2800  & \phn $\;$ 3.3 $\pm$ 0.03 & \phn 3.0  &     2670 $\pm$ 130      &  5, 9     \\
HD 30614   &  O9.5 Iae    &  29000  &  3.0 \phn  &     118  &  $0.33 \pm 0.04$  &  $3.21 \pm 0.3$  &     269 $\pm$ 21      &
        1200  &          34.7 $\pm$ 2.7  &     36.8  &     1240 $\pm$ 40  \phn &  3, 7, 9  \\
HD 36576   &  B2 IVe      &  26060  &  3.82  &     165  &  $0.45 \pm 0.06$  &  $3.28 \pm 0.4$  &     211 $\pm$ 14      &
    \phn 310  & \phn      7.0 $\pm$ 0.5  & \phn 7.3  & \phn 320 $\pm$ 10  \phn &  1, 2, 9 \\
HD 37737   &  O9.5 III    &  29800  &  3.95  &     182  &  $0.69 \pm 0.04$  &  $2.88 \pm 0.2$  & \phn 69 $\pm$  4 \phn &
        2300  &          17.1 $\pm$ 0.9  & \phn 7.2  &     1640 $\pm$ 660      &  4, 6, 8  \\
HD 52266   &  O9 V        &  31000  &  3.75  &     248  &  $0.31 \pm 0.01$  &  $3.32 \pm 0.4$  & \phn 66 $\pm$  5 \phn &
        1700  &          12.0 $\pm$ 0.9  &     13.4  &     1790 $\pm$ 90  \phn &  6, 7     \\
HD 52533   &  O9 V        &  32400  &  3.95  &     270  &  $0.21 \pm 0.03$  &  $3.17 \pm 0.5$  & \phn 42 $\pm$  1 \phn &
        2000  & \phn      9.0 $\pm$ 0.2  & \phn 7.9  &     1880 $\pm$ 120      &  6, 7     \\
HD 60848   &  O8 V:pevar  &  31600  &  4.20  &     163  &  $0.18 \pm 0.02$  &  $3.27 \pm 0.5$  & \phn 63 $\pm$  4 \phn &
        1900  &          12.8 $\pm$ 0.8  & \phn 7.2  &     1480 $\pm$ 420      &  6, 7     \\
HD 188001  &  O7.5 Ia     &  34080  &  3.36  & \phn 94  &  $0.32 \pm 0.05$  &  $3.18 \pm 0.2$  & \phn 99 $\pm$  5 \phn &
        3200  &          34.0 $\pm$ 1.6  &     20.8  &     2580 $\pm$ 620      &  6, 7     \\
HD 195592  &  O9.5 Ia     &  30460  &  3.19  &     114  &  $1.17 \pm 0.04$  &  $3.02 \pm 0.2$  &     220 $\pm$ 12      &
        1400  &          33.1 $\pm$ 1.8  &     22.1  &     1170 $\pm$ 230      &  6, 7     \\
\enddata
\tablerefs{
(1) \citealt{abt2002};
(2) \citealt{chauville2001}; 
(3) \citealt{crowther2006}; 
(4) \citealt{leitherer1988}; 
(5) \citealt{martin2006}; 
(6) \citealt{martins2005}; 
(7) \citealt{mason1998}; 
(8) \citealt{patriarchi2003}; 
(9) \citealt{schaller1992}; 
(10) \citealt{vansteenberg1988}. 
}
\end{deluxetable}

\clearpage
\begin{deluxetable}{lccccccc}
\rotate
\tablewidth{0pt}
\tablecaption{Peculiar Space Velocities\label{spacevel}}
\tablehead{
\colhead{ } &
\colhead{$\ell$} &
\colhead{$b$} &
\colhead{$\mu_\alpha \cos \delta$} &
\colhead{$\mu_\delta$} &
\colhead{$V_{t,pec}$} &
\colhead{$V_{r,pec}$} &
\colhead{$V_{pec}$} \\
\colhead{Star} &
\colhead{(deg)} &
\colhead{(deg)} &
\colhead{(mas yr$^{-1}$)} &
\colhead{(mas yr$^{-1}$)} &
\colhead{(\kms)} &
\colhead{(\kms)} &
\colhead{(\kms)} }
\startdata
HD 1976    &    118.68 &    $-10.63$ &  \phs    $13.70    \pm 0.25$     & \phn    $-4.40    \pm 0.42$     & \phn $4.8 \pm \phn 1.4$ & \phs   $-9.2 \pm 0.8$ &  $10.4 \pm \phn 1.6$ \\
HD 14633   &    140.78 &    $-18.20$ &  \phn    $-0.4\phn \pm 0.7$ \phn & \phn    $-7.8\phn \pm 1.0$\phn  &     $66.6 \pm     12.4$ &       $-24.1 \pm 1.7$ &  $70.8 \pm     12.5$ \\
HD 15137   &    137.46 & \phn$-7.58$ &  \phn\phs $0.67    \pm 0.40$     & \phn    $-5.08    \pm 0.73$     &     $57.0 \pm     11.0$ &       $-26.2 \pm 4.2$ &  $62.7 \pm     11.8$ \\
Feige 25   &    165.42 &    $-48.36$ &  \phn\phs $4.4\phn \pm 1.1$ \phn & \phn    $-0.3\phn \pm 1.2$\phn  &     $58.7 \pm     21.8$ & \phs   $33.2 \pm 4.3$ &  $67.4 \pm     22.2$ \\
HD 30614   &    144.07 &    $+14.04$ &  \phn\phs $0.49    \pm 0.18$     & \phn\phs $7.31    \pm 0.45$     &     $46.9 \pm \phn 3.2$ & \phs   $21.9 \pm 2.7$ &  $51.8 \pm \phn 4.2$ \\
HD 36576   &    187.39 & \phn$-7.84$ &  \phn\phs $0.15    \pm 1.20$     & \phn\phs $0.01    \pm 0.48$     & \phn $6.8 \pm \phn 1.9$ & \phs   $34.1 \pm 3.3$ &  $34.7 \pm \phn 3.8$ \\
HD 37737   &    173.46 & \phn$+3.24$ &  \phn\phs $0.8\phn \pm 0.6$ \phn & \phn    $-4.2\phn \pm 0.6$\phn  &     $25.5 \pm     13.9$ &       $-12.9 \pm 3.1$ &  $28.6 \pm     14.3$ \\
HD 52266   &    219.13 & \phn$-0.68$ &  \phn    $-0.89    \pm 0.78$     & \phn    $-0.70    \pm 0.61$     &     $18.9 \pm \phn 8.5$ & \phn   $-4.5 \pm 3.1$ &  $19.4 \pm \phn 9.0$ \\
HD 52533   &    216.85 & \phn$+0.80$ &  \phn    $-0.67    \pm 2.54$     & \phn    $-0.04    \pm 1.81$     &     $13.2 \pm     27.8$ & \phs   $45.1 \pm 2.0$ &  $47.0 \pm     27.9$ \\
HD 60848   &    202.51 &    $+17.52$ &  \phn    $-3.43    \pm 0.75$     & \phn    $-1.41    \pm 0.50$     &     $12.7 \pm \phn 7.4$ & \phn\phs$2.2 \pm 5.7$ &  $12.9 \pm \phn 9.4$ \\
HD 188001  & \phn56.48 & \phn$-4.33$ &  \phn    $-0.17    \pm 0.48$     &        $-10.45    \pm 0.34$     &     $81.5 \pm     22.0$ & \phn   $-2.3 \pm 4.3$ &  $81.5 \pm     22.4$ \\
HD 195592  & \phn82.36 & \phn$+2.96$ &  \phn    $-2.46    \pm 0.48$     & \phn\phs $1.62    \pm 0.51$     &     $37.2 \pm \phn 7.9$ &       $-20.1 \pm 2.1$ &  $42.3 \pm \phn 8.2$ 
\\
\enddata
\end{deluxetable}


\begin{thebibliography}{}
\bibitem[Abt et al.(1990)Abt, Gomez, \& Levy]{abt1990}
	 Abt, H. A., Gomez, A. E., \& Levy, S. G. 1990, \apjs, 74, 551
\bibitem[Abt et al.(2002)Abt, Levato, \& Grosso]{abt2002} 
	 Abt, H.~A., Levato, H., \& Grosso, M.\ 2002, \apj, 573, 359
\bibitem[Allen et al.(2006)Allen, Poveda, \& Hern\'andez-Alc\'antara]
	 {allen2006}
	 Allen, C., Poveda, A., \& Hern{\'a}ndez-Alc{\'a}ntara, A.  2006, 
	 Revista Mexicana de Astronomia y Astrofisica Conference Series, 
	 25, 13
\bibitem[Aslanov \& Barannikov(1992)]{aslanov1992} 
	 Aslanov, A.~A., \& Barannikov, A.~A. 1992, Soviet Astron.
	 Let., 18, 58
\bibitem[Aslanov et al.(1984)Aslanov, Kornilova, \& Cherepashchuk]{aslanov1984}
	 Aslanov, A.~A., Kornilova, L.~N., \& Cherepashchuk, A.~M. 1984, 
	 Soviet Astron. Let., 10, 278
\bibitem[Bally \& Zinnecker(2005)]{bally2005}
	 Bally, J., \& Zinnecker, H.  2005, \aj, 129, 2281
\bibitem[Beardsley(1969)]{beardsley1969} 
	 Beardsley, W.~R.\ 1969, Publ. of the Allegheny Observatory 
	 of the University of Pittsburgh, 8, 91
\bibitem[Berger \& Gies(2001)]{berger2001} 
	 Berger, D.~H., \& Gies, D.~R. 2001, \apj, 555, 364
\bibitem[Blaauw(1961)]{blaauw1961} 
	 Blaauw, A.\ 1961, \bain, 15, 26
\bibitem[Blaauw \& van Albada(1963)]{blaauw1963}
	 Blaauw, A., \& van Albada, T. S.  1963, \apj, 137, 791
\bibitem[Bohannan \& Garmany(1978)]{bohannan1978}
	 Bohannan, B., \& Garmany, C.~D.\ 1978, \apj, 223, 908
\bibitem[Bolton \& Rogers(1978)]{bolton1978}
	 Bolton, C. T., \& Rogers, G. L. 1978, \apj, 222, 234
\bibitem[Boyajian et al.(2005)]{boyajian2005}
	 Boyajian, T. S., Beaulieu, T. D., Gies, D. R., Huang, W., 
	 McSwain, M. V., Riddle, R. L., Wingert, D. W., \& De Becker, M.  
	 2005, \apj, 621, 978
\bibitem[Casares et al.(2005)]{casares2005}
	 Casares, J., Rib{\'o}, M., Ribas, I., Paredes, J.~M., 
	 Mart{\'{\i}}, J., \& Herrero, A.\ 2005, \mnras, 364, 899
\bibitem[Chauville et al.(2001)]{chauville2001} 
	 Chauville, J., Zorec, J., Ballereau, D., Morrell, N., Cidale, L., 
	 \& Garcia, A.\ 2001, \aap, 378, 861
\bibitem[Cohen et al.(2003)Cohen, Wheaton, \& Megeath]{cohen2003}
	 Cohen, M., Wheaton, W.~A., \& Megeath, S.~T.\ 2003, \aj, 126, 
	 1090
\bibitem[Colina et al.(1996)Colina, Bohlin, \& Castelli]{colina1996}
	 Colina, L., Bohlin, R. C., \& Castelli, F. 1996, HST Instr. Sci. 
	 Rep. CAL/SCS-008
\bibitem[Conti et al.(1977)Conti, Leep, \& Lorre]{conti1977}
	 Conti, P.~S., Leep, E.~M., \& Lorre, J.~J.\ 1977, \apj, 214, 759
\bibitem[Crowther et al.(2006)Crowther, Lennon, \& Walborn]{crowther2006} 
	 Crowther, P.~A., Lennon, D.~J., \& Walborn, N.~R. 2006, \aap, 
	 446, 279
\bibitem[Cutri et al.(2003)]{cutri2003}
         Cutri, R. M., et al. 2003,
         The 2MASS All-Sky Catalog of Point Sources
         (Pasadena: Univ. Mass. \& IPAC)
\bibitem[de Wit et al.(2004)]{dewit2004}
	 de Wit, W. J., Testi, L., Palla, F., Vanzi, L., \& Zinnecker, H.  
	 2004, \aap, 425, 937
\bibitem[De Zeeuw et al.(1999)]{dezeeuw1999}
	 De Zeeuw P.\ T., Hoogerwerf R., De Bruijne J.\ H.\ J., Brown A.\ 
	 G.\ A., \& Blaauw A. 1999, \aj, 117, 354
\bibitem[Docobo \& Costa(1986)]{docobo1986} 
	 Docobo, J.~A., \& Costa, J.~M.\ 1986, \apjs, 60, 945
\bibitem[Dray et al.(2005)]{dray2005} 
	 Dray, L.~M., Dale, J.~E., Beer, M.~E., Napiwotzki, R., \& King, 
	 A.~R.  2005, \mnras, 364, 59
\bibitem[Fullerton(1990)]{fullerton1990}
	 Fullerton, A. W. 1990, Ph.D. thesis, University of Toronto
\bibitem[Fullerton et al.(1996)Fullerton, Gies, \& Bolton]{fullerton1996} 
	 Fullerton, A.~W., Gies, D.~R., \& Bolton, C.~T.\ 1996, \apjs, 
	 103, 475
\bibitem[Garmany et al.(1980)Garmany, Conti, \& Massey]{garmany1980} 
	 Garmany, C.~D., Conti, P.~S., \& Massey, P.\ 1980, \apj, 242, 
	 1063
\bibitem[Gies \& Bolton(1986)]{gies1986}
	 Gies, D. R., \& Bolton, C. T.  1986, \apjs, 61, 419
\bibitem[Gray(1992)]{gray1992}
	 Gray, D. F. 1992, The observation and analysis of stellar
	 photospheres, 2nd ed. (Cambridge: Cambridge Univ. Press)
\bibitem[Gray(1998)]{gray1998}
	 Gray, R. O.  1998, \aj, 116, 482
\bibitem[Greenstein \& Sargent(1974)]{greenstein1974}
	 Greenstein, J.~L., \& Sargent, A.~I.\ 1974, \apjs, 28, 157
\bibitem[Gualandris et al.(2004)Gualandris, Portegies Zwart, \& 
	 Eggleton]{gualandris2004} 
	 Gualandris, A., Portegies Zwart, S., \& Eggleton, P.~P.\ 2004, 
	 \mnras, 350, 615
\bibitem[Hartkopf et al.(2000)]{hartkopf2000} 
	 Hartkopf, W.~I., et al. 2000, \aj, 119, 3084
\bibitem[Hauck \& Mermilliod(1998)]{hauck1998}
	 Hauck, B., \& Mermilliod, M.  1998, \aaps, 129, 431
\bibitem[H{\o}g et al.(2000)]{hog2000}
	 H{\o}g E., et al.  2000, \aap, 355, L27
\bibitem[Hoogerwerf et al.(2001)Hoogerwerf, de Bruijne, \& de 
	 Zeeuw]{hoogerwerf2001} 
	 Hoogerwerf, R., de Bruijne, J.~H.~J., \& de Zeeuw, P.~T.  2001, 
	 \aap, 365, 49
\bibitem[Humphreys(1978)]{humphreys1978}
	 Humphreys, R.~M.  1978, \apjs, 38, 309
\bibitem[Keenan \& Dufton(1983)]{keenan1983} 
	 Keenan, F.~P., \& Dufton, P.~L.\ 1983, \mnras, 205, 435
\bibitem[Kurucz(1994)]{kurucz1994}
	 Kurucz, R.\ L. 1994, Kurucz CD-ROM 19, Solar Abundance Model 
	 Atmospheres for 0, 1, 2, 4, 8 km/s (Cambridge: SAO)
\bibitem[Lanz \& Hubeny(2003)]{lanz2003} 
	 Lanz, T., \& Hubeny, I.\ 2003, \apjs, 146, 417
\bibitem[Leitherer(1988)]{leitherer1988}
	 Leitherer, C.  1988, \apj, 326, 356 
\bibitem[Leonard \& Duncan(1990)]{leonard1990} 
	 Leonard, P.~J.~T., \& Duncan, M.~J.\ 1990, \aj, 99, 608
\bibitem[Lyubimkov et al.(2004)Lyubimkov, Rostopchin, \& Lambert]
	 {lyubimkov2004} 
	 Lyubimkov, L.~S., Rostopchin, S.~I., \& Lambert, D.~L.\ 2004, 
	 \mnras, 351, 745 
\bibitem[Martin(2004)]{martin2004}
	 Martin, J.\ C.  2004, \aj, 128, 2474
\bibitem[Martin(2006)]{martin2006}
	 Martin, J.\ C.  2006, \aj, 131, 3047
\bibitem[Martins et al.(2005)Martins, Schaerer, \& Hillier]{martins2005}
	 Martins, F., Schaerer, D., \& Hillier, D.~J.\ 2005, \aap, 436, 
	 1049 
\bibitem[Mason et al.(1998)]{mason1998}
	 Mason, B.\ D., Gies, D.\ R., Hartkopf, W.\ I., Bagnuolo, W.\ G., 
	 ten Brummelaar, T., \& McAlister, H.\ A. 1998, \aj, 115, 821
\bibitem[Mason et al.(2001)]{mason2001} 
	 Mason, B.~D., Wycoff, G.~L., Hartkopf, W.~I., Douglass, G.~G., \& 
	 Worley, C.~E. 2001, \aj, 122, 3466
\bibitem[Mathias et al.(2001)]{mathias2001} 
	 Mathias, P., Aerts, C., Briquet, M., De Cat, P., Cuypers, J., Van 
	 Winckel, H., Flanders., \& Le Contel, J.~M.  2001, \aap, 379, 905
\bibitem[Mayer et al.(1994)]{mayer1994}
	 Mayer, P., Chochol, D.\ C., Hanna, M.~A.-M., \& Wolf, M.  1994, 
	 Contr.\ Astron.\ Obs.\ Skalnate Pleso, 24, 65
\bibitem[McSwain et al.(2004)]{mcswain2004}
	 McSwain, M.\ V., Gies, D.\ R., Huang, W., Wiita, P.\ J., Wingert, 
	 D.\ W., \& Kaper, L.  2004, \apj, 600, 927
\bibitem[McSwain \& Gies(2005)]{mcswain2005}
	 McSwain, M.~V., \& Gies, D.~R.  2005, \apjs, 161, 118
\bibitem[Meurs et al.(2005)Meurs, Fennell, \& Norci]{meurs2005}
	 Meurs, E.~J.~A., Fennell, G., \& Norci, L.\ 2005, \apj, 624, 307
\bibitem[Moffat et al.(1998)]{moffat1998} 
	 Moffat, A.~F.~J., et al.\ 1998, \aap, 331, 949
\bibitem[Morbey \& Brosterhus(1974)]{morbey1974}
         Morbey, C.\ L., \& Brosterhus, E.\ B. 1974, \pasp, 86, 455
\bibitem[Motch et al.(1998)]{motch1998}
	 Motch, C.\, et al.  1998, \aaps, 132, 341
\bibitem[Negueruela et al.(2004)Negueruela, Steele, \& Bernabeu]
	 {negueruela2004} 
	 Negueruela, I., Steele, I.~A., \& Bernabeu, G.\ 2004, 
	 Astronomische Nachrichten, 325, 749
\bibitem[Noriega-Crespo et al.(1997)Noriega-Crespo, van Buren, \& 
	 Dgani]{noriega1997} 
	 Noriega-Crespo, A., van Buren, D., \& Dgani, R.\ 1997, \aj, 113, 
	 780
\bibitem[Ochsenbein et al.(2000)Ochsenbein, Bauer, \& Marcout]
	 {ochsenbein2000} 
	 Ochsenbein, F., Bauer, P., \& Marcout, J.\ 2000, \aaps, 143, 23
\bibitem[Patriarchi et al.(2003)Patriarchi, Morbidelli, \& 
	 Perinotto]{patriarchi2003} 
	 Patriarchi, P., Morbidelli, L., \& Perinotto, M.\ 2003, \aap, 
	 410, 905
\bibitem[Perryman et al.(1997)]{perryman1997} 
	 Perryman, M. A. C., et al. 1997, \aap, 323, L49
\bibitem[Philp et al.(1996)]{philp1996}
	 Philp, C.~J., Evans, C.~R., Leonard, P.~J.~T., \& Frail, D.~A.\ 
	 1996, \aj, 111, 1220
\bibitem[Plaskett(1924)]{plaskett1924} 
	 Plaskett, J.~S.\ 1924, Publ. Dominion Astrophys. Obs. Victoria, 
	 2, 285
\bibitem[Portegies Zwart(2000)]{portegieszwart2000}
	 Portegies Zwart, S.~F.\ 2000, \apj, 544, 437
\bibitem[Poveda et al.(1967)Poveda, Ruiz, \& Allen]{poveda1967} 
	 Poveda, A., Ruiz, J., \& Allen, C.  1967, Bol. Obs. Tonantzintla 
	 Tacubaya, 4, 86
\bibitem[Rib{\'o} et al.(2002)]{ribo2002} 
	 Rib{\'o}, M., Paredes, J.~M., Romero, G.~E., Benaglia, P., 
	 Mart{\'{\i}}, J., Fors, O., \& Garc{\'{\i}}a-S{\'a}nchez, J. 
	 2002, \aap, 384, 954
\bibitem[Roberts et al.(1987)Roberts, Leh\'{a}r, \& Dreher]{roberts1987}
         Roberts, D.\ H., Leh\'{a}r, J., \& Dreher, J.\ W. 1987, \aj, 93,
         968
\bibitem[Rzayev \& Panchuk(2004)]{rzayev2004} 
	 Rzayev, A.~K., \& Panchuk, V.~E.\ 2004, Astr.\ Let., 30, 332
\bibitem[Sayer et al.(1996)Sayer, Nice, \& Kaspi]{sayer1996}
	 Sayer, R.~W., Nice, D.~J., \& Kaspi, V.~M.\ 1996, \apj, 461, 357
\bibitem[Schaller et al.(1992)]{schaller1992} 
	 Schaller, G., Schaerer, D., Meynet, G., \& Maeder, A.\ 1992, 
	 \aaps, 96, 269
\bibitem[Stickland \& Lloyd(2001)]{stickland2001}
	 Stickland, D.\ J., \& Lloyd, C.  2001, The Observatory, 121, 1
\bibitem[Stone(1982)]{stone1982}
	 Stone, R.~C.\ 1982, \apj, 261, 208
\bibitem[Thompson et al.(1978)]{thompson1978}
	 Thompson, G.\ I., Nandy, K., Jamar, C., Monfils, A., Houziaux, 
	 L., Carnochan, D.\ J., \& Wilson, R. 1978, Catalogue of stellar 
	 ultraviolet fluxes: A compilation of absolute stellar fluxes 
	 measured by the Sky Survey Telescope (S2/68) aboard the ESRO 
	 satellite TD-1, The Science Research Council, U.\ K.\ (VizieR 
	 No.\ II/59B)
\bibitem[Underhill(1995)]{underhill1995}
	 Underhill, A.\ B. 1995, \apjs, 100, 433
\bibitem[Underhill \& Matthews(1995)]{underhill1995b}
	 Underhill, A.\ B., \& Matthews, J.\ M.  1995, \pasp, 107, 513
\bibitem[Urban et al.(2004)Urban, Zacharias, \& Wycoff]{urban2004} 
	 Urban, S.~E., Zacharias, N., \& Wycoff, G.~L.\ 2004, U.\ 
	 S.\ Naval Observatory, Washington, D.\ C.\ (VizieR No.\ I/294)
\bibitem[van Steenberg \& Shull(1988)]{vansteenberg1988}
	 van Steenberg, M.~E., \& Shull, J.~M.  1988, \apjs, 67, 225
\bibitem[Voges et al.(2000)]{voges2000}
	 Voges, W., et al.  2000, \iaucirc, 7432, 1
\bibitem[Walborn(1971)]{walborn1971}
	 Walborn, N.\ R.  1971, \apj, 164, 67
\bibitem[Wegner(2000)]{wegner2000} 
	 Wegner, W.\ 2000, \mnras, 319, 771
\bibitem[Wesselius et al.(1982)]{wesselius1982}
	 Wesselius P.\ R., van Duinen R.\ J., de Jonge A.\ R.\ W., Aalders 
	 J.\ W.\ G., Luinge W., \& Wildeman K.\ J.  1982, \aaps, 49, 427
\bibitem[Zacharias et al.(2004)]{zacharias2004} 
	 Zacharias, N., Urban, S.~E., Zacharias, M.~I., Wycoff, G.~L., 
	 Hall, D.~M., Monet, D.~G., \& Rafferty, T.~J.\ 2004, \aj, 127, 3043
\bibitem[Zejnalov \& Musaev(1986)]{zejnalov1986} 
	 Zejnalov, S.~K., \& Musaev, F.~A.\ 1986, Pisma Astronomicheskii 
	 Zhurnal, 12, 304
\bibitem[Zorec et al.(2005)Zorec, Fr\'emat, \& Cidale]{zorec2005} 
	 Zorec, J., Fr{\'e}mat, Y., \& Cidale, L.\ 2005, \aap, 441, 235 
\bibitem[Zwicky(1957)]{zwicky1957}
	 Zwicky, F.\ 1957, Morphological Astronomy (Berlin: Springer)
\end{thebibliography}
\end{document}